\documentclass{article}
\usepackage{spconf,amsmath,graphicx}
\usepackage{booktabs}
\usepackage{hyperref}
\usepackage[export]{adjustbox}

\def\x{{\mathbf x}}

\title{One-Shot Acoustic Matching Of Audio Signals --
\\ Learning to Hear Music In Any Room/ Concert Hall
}
\name{Prateek Verma, Chris Chafe, Jonathan Berger}
\address{Stanford University
\\prateekv@stanford.edu}

\vspace{-2cm}
\begin{document}

\maketitle
\begin{abstract}
\vspace{-0.2cm}
The acoustic space in which a sound is created and heard plays an essential role in how that sound is perceived by affording a unique sense of \textit{presence}. Every sound we hear results from successive convolution operations intrinsic to the sound source and external factors such as microphone characteristics and room impulse responses. Typically, researchers use an excitation such as a pistol shot or balloon pop as an impulse signal with which an auralization can be created. The room "impulse" responses convolved with the signal of interest can transform the input sound into the sound played in the acoustic space of interest. Here we propose a novel architecture that can transform a sound of interest into any other acoustic space(room or hall) of interest by using arbitrary audio recorded as a proxy for a balloon pop.
The architecture is grounded in simple signal processing ideas to learn residual signals from a learned acoustic signature and the input signal. Our framework allows a neural network to adjust gains of every point in the time-frequency representation, giving sound qualitative and quantitative results.
\end{abstract}
\vspace{-0.3cm}
\section{Introduction and Related Work}
\vspace{-0.4cm}
Humans interact with a rich palette of sounds \cite{gemmeke2017audio} in a wide range of acoustic environments. We experience music not only in individual settings, such as headphones and home speakers but also in highly reverberant spaces, such as theatres and concert halls. Listening to live music performed in these acoustic spaces often adds enjoyment and richness to the experience \cite{rossing2004principles}. The design of these acoustic spaces shaped the input signals' timbre and volume dynamics, sometimes precisely according to the genre and type of music performed \cite{bagenal1930bach}. Detailed studies of cave acoustics inhabited by paleolithic humans suggest that the reverberant qualities of these caves might have functioned as the first auditoriums in which humans experienced sound \cite{fazenda2017cave}. It is conceivable that as early humans found safety and comfort inhabiting caves, they also found comfort in their dwelling's rich acoustical environment \cite{shipton201878}. One can also hypothesize that human evolution has adapted to reverberant sounds, contrasted by discomfort with anechoic sounds, and reverb adds pleasantness to the sounds. As shown in \cite{maconie2010chapter}, musicians/listeners dislike experiencing and playing music in environments void of natural reflections, aka anechoic chambers. Neural architectures have recently revolutionized the field of audio signal processing. With the advent of Transformer based architectures \cite{vaswani2017attention}, there has been a pivot on approaching almost all problems in areas such as computer vision \cite{dosovitskiy2020image}, NLP \cite{vaswani2017attention,wei2021finetuned} and audio \cite{verma2021audio,verma2021generative,dhariwal2020jukebox}, with powerful attention based architectures. This work touches on ways to derive audio embeddings: do a conditioned transformation based on these embeddings. Audio embeddings have been powerful to aid in a variety of applications such as ASR, Audio Understanding \cite{Chung2018-Speech2Vec}, \cite{haque2019audio}, conditional audio synthesis \cite{haque2019audio,skerry2018towards} as well as transformation \cite{oord2017neural, verma2018neural}. These latent vectors can also be used for summarizing the contents of the audio signal: then use a head similar to \cite{chen2020simple, wang2021multi} for classification purposes. There has been a recent surge of interest in transforming audio into a desired acoustic space both in academia and industry. \cite{su2020acoustic} proposed a way of embedding impulse responses to conditionally them and then to guide a generative auto-regressive architecture resembling wavenet \cite{oord2016wavenet}. 
\begin{figure}
    \includegraphics[width=0.5\textwidth,left]{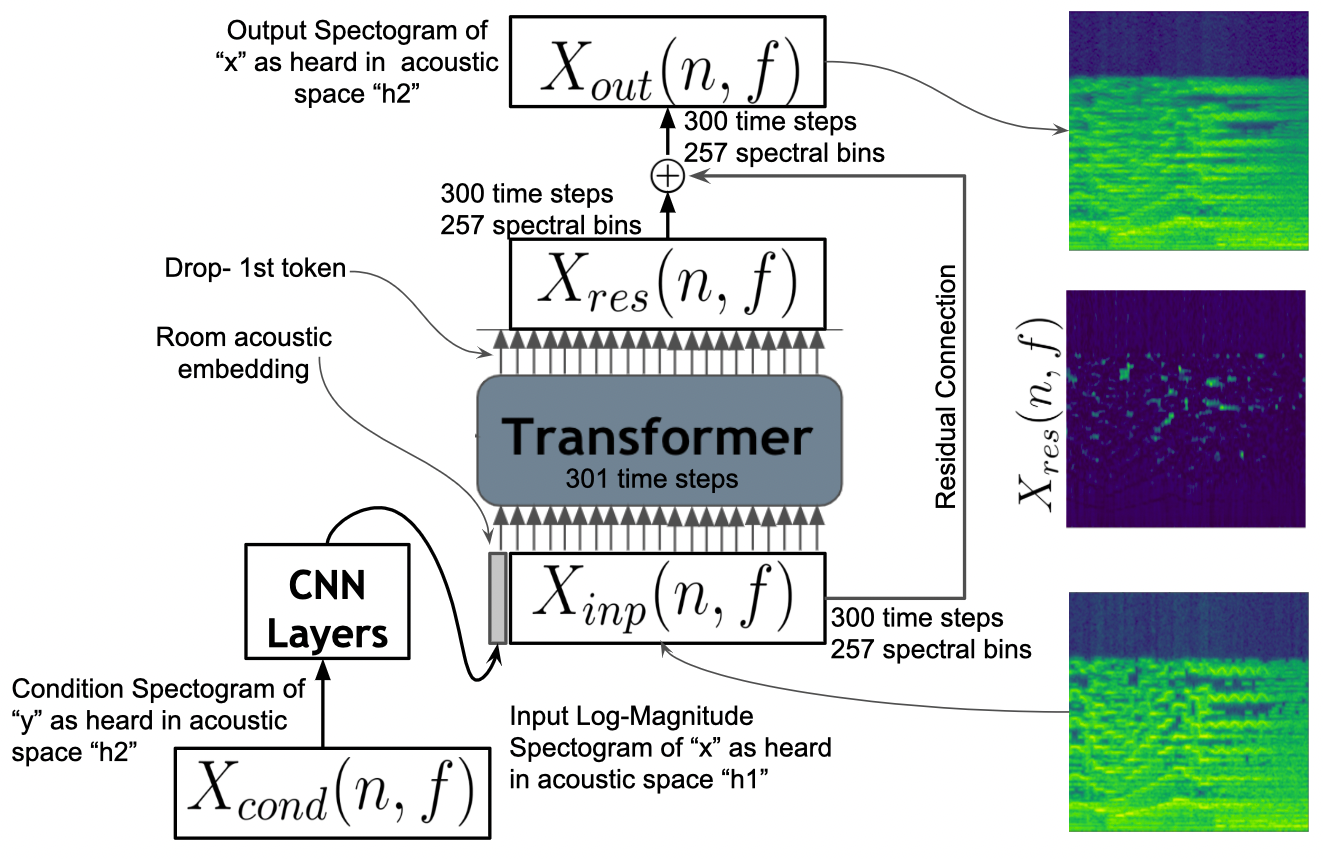}
    \caption{A single snippet of sound "y" is used as a proxy for balloon pop, typically used to measure room acoustics. Transformer architecture is trained to do a conditional audio transformation of "x" as heard in acoustic space "h1" to "h2" by using the learned residual. With a trained model and conditioning audio, we can transform the audio input now to the desired space of interest.}
    \label{fig:img1}
\end{figure}
This process is slow due to predicting auto-regressive samples one at a time. \cite{Singh_2021_ICCV} proposed a way of directly generating room impulse responses from a 2-D image. Since measuring and recording impulse response via traditional methods, such as a balloon pop, is time-consuming and challenging to scale, the authors proposed this workaround. Similar methods extending to a 3-D environment to estimate impulse responses directly from 2-D/3-D data have been proposed recently \cite{majumder2022few}. \cite{chen2022visual} used an image of a target environment to compute an embedding and a waveform of source audio to re-synthesize the audio conditionally using an embedding. It used a complex adversarial-based formulation. With these methods, the data collection and building of the desired dataset are complex. Moreover, even within an image of a single acoustic environment, the impulse response varies significantly (closer to walls vs. windows vs. center), which audio signals capture. The visual data beyond its field of view cannot capture the complete geometry that the audio signals can capture. The contributions of this work are as follows: 1. We show (for the first time, to our knowledge) condition-based acoustic transfer using audio signals as a condition in a non-autoregressive setting. 2. We use a simple architecture inspired by signal processing properties of convolution to learn residual signals, which directly learn a residual signal instead of a waveform or a spectrogram. 3. We use a novel min-max loss function that circumvents some drawbacks of regression-based loss functions such as euclidean loss and mean absolute error. 4. In this work, we showcase for the first time in audio and signal processing literature a robust Transformer based framework that can map spectrogram to spectrogram based on the conditioning of audio signals. \vspace{-0.5cm}
\section{Dataset} \vspace{-0.4cm}
Since, to the best of our knowledge, no such work existed for music signals, we built the framework ourselves. With the ubiquitous nature of transformer architectures, we expect our approach to holding for other kinds of audio signals. Since large-scale anechoic data is unavailable for a large variety of music except for subgenres,  \cite{d2020anechoic}, we use a mixture of mixture approach. The initial audio signals are real-world recordings of composers like Mozart, Vivaldi, and Beethoven encompassing a range of instruments such as piano, violin, and symphonic. For reproducibility purposes, we have shared the audio files at the following URL: http://ai.stanford.edu/~prateekv/testurl.txt. These files mimic a variety of playing conditions. Let us assume that for a concert hall, $h^{i}_{ir}[n]$ is the room impulse response present already from the location in which the audio gets recorded. Considering the microphone acoustics, intrinsic room acoustics, the recording compression, and the instrument sound production system, $x_{i}[n]$ was the input signal. The input dataset is a collection of signals $x[n]$ each of which is produced by convolving the signal $x_{i}[n]$ with $h^{i}_{ir}[n]$. We further convolve it with different conditioning/input IRs. \vspace{-0.4cm}
\subsection{Impulse Response}
\vspace{-0.2cm}
We use a dataset of synthetic impulse responses to simulate the effects of various acoustic spaces (concert halls and rooms). Techniques like balloon pop, sweeping sinusoids, or as proposed by \cite{abel2010estimating} are used in modeling the room acoustics. However, these are non-trivial to start with and are difficult to scale. Therefore, we started with \cite{ko2017study} as the primary source of synthetic impulse responses. To the best of our knowledge, this was the largest-sized dataset in terms of the number of unique impulse response measurements known to us. In addition, there are 200 rooms, each uniformly distributed between small, medium, and large, with various room sizes, absorption coefficients, heights, and locations inside every room, yielding quite a diverse set of IRs.
\subsection{Combining the two}
\vspace{-0.2cm}
Now given a collection of thousands of small patches of these signals $x[n]$ and room impulses $h^{pr}[n]$, we generate the training data. The goal of our work is to give an acoustic signal played in an acoustic space to transform it into another acoustic space. The current work uses a simple convolution to mimic how audio would sound in an acoustic space. Given a signal of interest, $x[n]$, it would sound in a room r at point p, as $y[n] = x[n] \ast h^{pr}[n] $. Realistic reverberation using signal processing algorithms is a field in itself, and \cite{valimaki2016more} describes a detailed study. We carry out a similar approach to generate the conditioning audio spectrogram.
\section{Methodology}
\vspace{-0.2cm}
This section describes the components and the architectural choice of the current work. Broadly the goal is to transform the input audio spectra $X_{a}^{i}$ to that of $X_{a}^{j}$, with conditioning audio spectra $X_{b}^{j}$, where $X_{a,b}^{i,j}$ is audio $a,b$ played in acoustic space (impulse response) $i,j$. 
\vspace{-0.25cm}
\subsection{Need For Residual Learning}
\vspace{-0.2cm}
We aim to transform the input content audio into a different acoustic space. We would presume that to play the audio in an acoustic space, whose room impulse response is $h_{d}[n]$ at a particular location in space, then the desired output audio $y[n]$ of interest is, 
$y[n] = x_{i}[n] \ast h_{d}[n]$. Instead of having access to $h_{d}[n]$, we have the conditioning audio as input. Therefore, our architecture should extract a latent space for $h_{d}[n]$ directly from the conditioning audio or derive a similar latent space. Thus given a input audio $x_{i}[n]$ and a latent space embedding $h_{d}[n]$, we should be able to extract $y[n]$. We explore working in the frequency domain because a property of convolution is that multiplication in the frequency domain is convolution in the time domain. Let us assume $X_{i}(f,n)$ is the STFT representation of the input signal, and $H_{d}(f,n)$ is that of the impulse response. Then, the output STFT is,
$Y(f,n) = X_{i}(f,n) \odot H_{d}(f,n) $. Now, working in the log domain, 
\begin{equation}
     log{Y(f,n)} = log X_{i}(f,n) + log H_{d}(f,n) 
\end{equation}

Therefore, for log-magnitude spectrograms, we need to add a signal or its \textbf{\textit{residual}} component. This framework falls into the realm of residual learning \cite{he2016deep}, and neural architectures are good at learning residuals, yielding better convergence and improved performance \cite{he2020resnet}. Thus for our conditioned generation of audio signals, to generate the desired transformation as seen from Eqn (1), we take in an audio spectrogram and devise a neural architecture that can learn a residual signal which we add back to the input, to give the desired transformation.
\vspace{-0.4cm}
\subsection{Deriving Acoustic Signature \& the Setup}
\vspace{-0.2cm}
Many papers have explored neural architectures that can conditionally do the desired transformation, the closest to our work being \cite{haque2018conditional}. A primary architecture uses the commands as latent embeddings to do the desired transformation. We use a  architecture similar to \cite{su2020acoustic} to extract acoustic space embeddings. The final layer passes through a global average pooling layer to get an embedding that acts as a condition to a Transformer architecture. Our primary transformational architecture is a Transformer, having attention mechanisms capable of understanding long-term dependencies. It also allows us to do condition-based transformation as seen in \cite{verma2021generative} and \cite{keskar2019ctrl}. We devise a Transformer architecture that can map the input of dimensions $257\x 300$ to the same dimensional output, where 300 corresponds to the time steps in 3s, with each timestep a log-magnitude Fourier transform a slice of dimension 257. Sinusoidal positional encodings are added as in \cite{vaswani2017attention}. The embedding size is 257, with the feed-forward layers being 512 and 8 head attention heads for three layers. Conditional embedding size 257 is concatenated to position 0, thus totaling 301 tokens in the first and the output of the last layer of the Transformer. This ensures that Transformer can attend to this token as and when needed in whatever hierarchy it chooses. We slice the output of the last layer of the Transformer to have only 300 tokens, as desired. The attention mechanism ensures that the correct log-magnitude Fourier slices are present at suitable locations. The output of the last layer of the Transformer is called a residual signal. According to the conditioning and input audio, a residual signal can increase/decrease the gain of each time-frequency point depending on the input signal. Adding this residual signal to the input log-magnitude spectrogram gives us the desired transformation. 
\vspace{-0.4cm}
\subsection{The Loss Function}
\vspace{-0.2cm}In continuous value regression problems, mean squared error(MSE) or mean absolute error(MAE) is minimized as an error criterion to be minimized. However, both of these metrics are scale-dependent, i.e., the larger the value larger the error. Our experiments also found that minimizing MAE/MSE first focused on higher harmonics(typically having larger values) than lower-order harmonics. However, all the harmonics are equally crucial for the perception of music. This can be one of the main reasons across the literature; $L1$ /$L2$ loss is not used with adversarial loss function, which is difficult to tune in a multi-criterion setup. We propose a novel min-max loss formulation inspired by literature in classic optimization. Let us define two spectrograms $X_{p}$ and $X_{t}$ denoting predicted/target spectrograms, each consisting of $f$ frequency points (257 in our case) and $n$ time points (300 in our case). We define min-max loss formulation as a single scalar measuring the distance between the desired output and the predicted output we choose to minimize. Mathematically,
\vspace{-0.2cm}
$$\mathcal{L} (X_{t},X_{p}) = \sum_{\forall k \in [1,n]} \max\limits_{f}  {|(X_{t}(k,f)- X_{p}(k,f))|} $$
This ensures that the prediction is not biased against the largest values within a spectrogram slice. To ensure we remove implicit volume-dependent biases, we randomly scale the input volume with random maximum values chosen between [0,1]. Finally, we sum the maximum deviation across a single time slice across all the time points to get the desired scalar telling how far apart the prediction is for the neural architecture to minimize. 
Our experiments produced much richer audio, particularly in the higher harmonics, compared to the $L1$ and $L2$ loss criterion. During inference, we use Griffin-Lim reconstruction \cite{griffin1984signal} to go from the generated spectrogram to the actual waveform. Furthermore, we used data augmentation techniques for the input signal. In addition to the techniques to modify the input signals, \cite{mcfee2015software} dropout rates were tuned to close the training-validation loss gap. As noted in \cite{hershey2017cnn}, using a large amount of training data is a good regularizer. We train all the architectures for about 100 epochs. Adam optimizer with a learning rate of 2*1e-4  was reduced till 1e-6 whenever the loss started plateauing. We had a total of 300,000 3s patchs randmoly sample from the training audio data of roughly 20 hours with random room impulse responses for our training dataset.

\section{Results and Discussion}
\vspace{-0.2cm}To demonstrate the results of our work, we share some audio examples, perform listening experiments, and devise quantitative experiments demonstrating evidence of the transformation. To share the results, we have put up a small subset of the transformations on a webpage. 
This webpage contains different audio examples, conditioning audio, the input audio transformed to the acoustic space of interest, the target audio, and our predicted audio. \newline
http://ai.stanford.edu/\(\sim \)prateekv/IRtransfer.html
\vspace{-0.2cm}
\subsection{Quantitative Results}
\vspace{-0.2cm}
We devise an experiment to show quantitatively the transformation of interest in our paper. Similar to the listening test, we propose a typical convolutional-based architecture to score how close two audio signals are in terms of acoustic space. E.g., if we play two audio signals in the same acoustic space, then the neural architecture (CNN network henceforth) should give a score of 0, whereas if they are playing in a different acoustic space, then it should give us a score of 1. We train a convolutional architecture in a siamese setup, irrespective of the content of the input signal (e.g., the two audio signals can have different instruments/composers or a mixture of two). First, we devise positive pairs (same acoustic space) as follows: i) For half of the positive sample pairs, we take the same audio with the same impulse response. Then, the audio is randomly scaled, and augmented copies of the input spectrogram are assigned a score of 0. For the second half of the positive pairs, we take in the same audio content: randomly scale the amplitudes but convolve it with different IRs to get two signals, $audio_{1} \ast imp_{1}$  and $audio_{1} \ast imp_{2}$. ii) For creating negative pairs (different acoustic spaces), for the first half, we randomly sample an audio signal $audio_{1}$ and convolve it with two differently sampled impulse signals $imp_{1}$ and $imp_{2}$ to get $audio_{1} \ast imp_{1}$, and $audio_{1} \ast imp_{2}$. For the second half of the negative pairs, we randomly sample two different audio signals as well as the impulse responses and compute  $audio_{1} \ast imp_{1}$, and $audio_{2} \ast imp_{2}$. The validation and the test set do not contain overlapping content or IRs. The convolutional architecture takes in the spectrogram representation of two audio signals $audio_{1}$ and $audio_{2}$ and assigns a score of 0 for the same acoustic space and 1 for a different acoustic space. The log-spectrogram computed with 10ms hop, with \cite{mcfee2015librosa} libraries, with 512 point FFT. We used typical data-augmentation strategies such as volume/amplitude scaling, random flips, random cutout, and jittering, as described in \cite{mcfee2015software} with every audio signal. This data-augmented and passed onto a convolutional encoder. We use an Efficient-Net B0 \cite{tan2019efficientnet} architecture that can take 3s of audio spectrogram representation of size (257x300) and map it to a latent embedding of size 128 dim. With two input audio spectrograms, we get two vectors $emb1$ and $emb2$ of 128-dims. We do not have a large classification head to force encoders to extract good embeddings. We take the embeddings, subtract them, and use a linear head of dimension 256 followed by an output layer of size 2. Cross entropy loss minimizes the prediction error between the target and the predicted output. We obtained an accuracy of about 94 \%, which in itself is quite remarkable. It now gives us a proximity score of two pieces played in the same or different rooms. We compare the scores for both the input signal and the conditioning signal (before transformation) and the input signal and the predicted transformation. The average score across the validation set of about 2000 snippets across several rooms decreases from 0.95 to 0.6. We achieve good scores from the listening experiment, yet we do not quite reach a small score with a neural architecture doing the same task. One way of looking at this is that the neural evaluator is a "very picky and idealistic" distance function, which can pick up subtle differences for any two audio spectrograms anywhere in the present to be present in a different/similar acoustic space. However, this idealistic evaluator still indicates the transformed audio moving closer to the acoustic space of interest.
\vspace{-0.45cm}
\subsection{Listening Experiment Results}
\vspace{-0.3cm}
For conducting the listening experiments, we ask users two questions.  i) 1. Given the desired output, what is the quality of the output ii) Are the input and the transformed audio similar/different? In other words, given the transformed output, is it closer to the input or the desired target? For 100 randomized ratings, we use human listeners to rate them via a webpage to validate the results. In the experiment, for evaluating the quality, we ask the users to rate both the actual ground truth target and what was predicted by our architecture separately on a mean opinion score scale (MOS) scale between 1 to 5, with 1 being poor quality. We obtain a 3.6 MOS score for our predicted audio, with the ground truth being 4.1. Finally, we ask users to rate the transformation if the predicted output is close to that of the input or target audio. We get the human ratings correct about 76\% of the times, showing that the transformed output and the ground truth are being played in the same space, validating our experiments.
\vspace{-0.2cm}
\section{Conclusion and Future Work} 
\vspace{-0.3cm}
We have demonstrated a pipeline that can do an acoustic transfer of input signal to an acoustic space of interest irrespective of the conditioning's content (e.g., piece, instrument, composer). A neural architecture with ideas grounded in signal processing using a novel min-max loss function is described. A CNN architecture to extract acoustic signatures is used to condition a transformer architecture. A generation transformer uses the learned acoustic signature and input signal to generate a conditionally dependent output spectrogram. Subjective listening tests show how human raters validate the claims of acoustic transfer to a different reverberant environment. Finally, we also develop neural architecture-based scores that objectively confirm the listening test of human raters. It will be interesting to extend this work to speech signals and jointly optimize our scoring mechanism with a generator. Finally, as shown, conditional generation using a robust attention-based model is a compelling idea and has applications far beyond the current work.
\section{Acknowledgement}
We thank the Institute of Human-Centered AI at Stanford University (Stanford HAI) for supporting this work through a generous Google Cloud computing grant for the academic year 2021-22 to carry out computational experiments. We thank both the HAI as well as Google for the initiative. Research was, in part, supported by the Templeton Religion Trust's \textit{Art Seeking Understanding} initiative. We would also like to thank Prof. Preeti Rao and students of DAPLAB and EE 679 at IIT Bombay and students of Stanford University for taking the listening experiments

\bibliographystyle{IEEEbib}
\bibliography{bib}
\end{document}